\documentclass[a4paper,12pt,one column]{article}
\usepackage{graphicx}
\usepackage{colordvi}
\usepackage{amssymb}
\usepackage{latexsym}
\usepackage{hyperref}
\begin{document}
\title{ Vacuum Energy Accretion and Primordial Black Holes in Brans-Dicke theory}
\author{D. Dwivedee$^{*}$, B. Nayak$^{\dag}$ and L. P. Singh$^{\ddag}$ \\
Department of Physics, Utkal University,
Bhubaneswar 751004, India. \\
$^{*}$debabrata@iopb.res.in \\ $^{\dag}$bibeka@iopb.res.in \\ $^{\ddag}$lambodar$\_$uu@yahoo.co.in \\}
\date{ }
\maketitle
\begin{abstract}
In this paper, we investigate the evolution of primordial black hole (PBH) in vacuum-dominated era within Brans-Dicke cosmology. We consider the accretion of vacuum energy by PBHs and found that vacuum energy accretion efficiency should be less than 0.61. We also study the evaporation of PBHs where we conclude that larger the value of accretion efficiency longer live the PBHs. We also find that PBHs evaporate at a quicker rate in Brans-Dicke theory compared with Standard Cosmology \cite{nj}.
\end{abstract}

\section{INTRODUCTION}
Einstein's General Theory of Relativity (GTR) \cite{ein} is based on a pure tensor
theory of gravity where gravitational constant is taken as a
time-independent quantity. But Brans-Dicke (BD) theory \cite{bdt} is a scalar-tensor
theory of gravity where the gravitational constant is a time-dependent
quantity. BD theory is the simplest extension over GTR through the
introduction of a time-dependent scalar field $\phi(t)$ as $G(t)$ $\sim $
$\phi^{-1}(t)$, where the scalar field $\phi(t)$ couples to gravity with a
coupling parameter $\omega$ known as the BD parameter. Interestingly BD
theory can be transformed to GTR in the limit $\omega$ $\rightarrow$
$\infty$ \cite{bam1, bam2}. Solar system observations require $\omega > 10^4$ \cite{nature}. Also BD type model can be regarded as the lowest limit of
Kaluza-Klein and String theories \cite{asm1, asm2, asm3}. Again BD theory explains many cosmological
phenomena such as inflation \cite{johri, la}, early and late time behaviour of the
universe \cite{sahoo1, sahoo2}, cosmic acceleration and structure formation \cite{bermar}, coincidence
problem \cite{nayak} and problems relating to black holes \cite{ijtp}.

                           Primordial Black Holes (PBHs) are those black
holes which are formed in the early universe through variety of mechanisms
such as inflation \cite{cgl, kmz}, initial inhomogeneities \cite{carr, swh}, phase transition and
critical phenomena in gravitational collapse \cite{khopol1, khopol2, khopol3, khopol4, jedam1, jedam2, jedam3}, bubble collision \cite{kss} or the
decay of cosmic loops \cite{polzem1, polzem2}. A comparision of cosmological density of the
universe with the density associated with a black hole, at any time after
BigBang, shows that formation mass of PBH would have same order as that of horizon mass. Thus PBH could span wide mass range starting from Planck mass
$10^{-5} g$ to more than $ 10^{15} g$. From Hawking's point of view black
holes emit thermal radiation quantum mechanically \cite{hawk}. So black holes will
evaporate depending upon their formation masses. Smaller the mass of
PBHS, quicker they evaporate. As density of a black hole varies inversely
with its mass, high density which is possible in the early universe, is
required to form lighter black holes. So PBHs are the only black holes
whose masses could be so small that they became able to evaporate
completely by the present epoch through Hawking evaporation \cite{hawk}. Early
evaporating PBHs could account for baryogenesis \cite{bckl, mds1, mds2}, in the universe. On the
other hand, Longer lived PBHs could act as seeds for structure formation \cite{mor1, mor2, mor3, mor4, mor5} 
and could also form a significant component of the dark matter \cite{blais1, blais2, blais3, blais4}. Again
 in refs \cite{nsmprd1, nsmprd2, pramana}, it is concluded  that PBHs
could take comparatively more time to evaporate due to accretion of
radiation which makes them long lived. 

Accordingly the standard picture of cosmology, the universe is
radiation-dominated in the very beginning of its evolution and now it is
matter-dominated. This gives the decelerated expansion of universe through
its evolution. But the observations of distant measurements to type Ia
Supernovae (SNIa) \cite{apj} indicate that the expansion of universe is
accelerating in the present epoch and that two-third of the critical
energy density exists in a dark energy component with a large
negative pressure and of unknown composition. The simplest
possibility of dark energy is vacuum energy. Recent data shows that dark energy 
occupies $68.3\%$ of universe and $31.7\%$ is occupied by some other
matter. SNIa observations also provide the evidence of a decelerated
universe in the recent past with transition from decelerated to
accelerated occuring at redshift $z_{q=0}\sim0.5$ \cite{mst}. So the vacuum energy
should be started from $z_{q=0}\sim 0.5$ i.e. $t_{q=0} \sim 0.5 t_0$. The
equation of state parameter $\gamma=-1$ is the most acceptable candidate
for dark energy. 

In this work, we examine the evolution of PBHs only in
vacuum-dominated era using BD theory. In our study, we consider vacuum energy accretion by PBHs and its effect on their evaporation. We also present a comparision between our results and the corresponding results of Standard Cosmology \cite{nj}.

\section{Solutions of scale factor $a(t)$ and gravitational constant
$G(t)$ in vacuum dominated era}

For a spatially flat $(k=0)$ Friedmann-Robertson-Walker universe with scale
factor $`a(t)'$ the first Friedmann equation using Brans-Dicke theory in
vacuum dominated era is
\begin{eqnarray} 
\frac{{\dot{a}}^{2}}{a^{2}}+\frac{\dot{a}}{a}\frac{\dot{\phi}}{\phi}
-\frac{\omega}{6}\frac{{\dot{\phi}}^{2}}{{\phi}^{2}}=\frac{8\pi
\rho_{v}}{3\phi}
\end{eqnarray}
where $\rho_v $ is the vacuum energy density 	

The energy conservation equation is given by

\begin{equation}
\dot{\rho}+3(\gamma + 1) H\rho = 0
\end{equation}

$\gamma$ is  the equation of state parameter and for vacuum energy $\gamma = -1$, $ H={\dot{a}}/a$ is the Hubble parameter.
So equation (2) gives vacuum energy density $\rho_{v} $ is a constant.

From our previous paper \cite{ijmpd}, we have obtained the time dependent gravitational constant for vacuum dominated era by matching the time dimension of each term of Friedmann equation (1) as 

\begin{equation}
G(t)=G_0\Big(\frac{t_0}{t}\Big)^2
\end{equation}
where $G_0$ is the present value of the gravitational constant and $t_0$ is the present age of the universe.

Using equation (3), the above Friedmann equation can be expressed as

\begin{eqnarray}
\frac{{\dot{a}}^2}{a^2}+\frac{2}{t}\frac{\dot{a}}{a} - \Big(\frac{8\pi G_0}{3} \frac{{t_0}^2}{t^2}\rho_v +\frac{2}{3}\frac{\omega}{t^2} \Big)=0
\end{eqnarray}

The solution of this equation gives

\begin{eqnarray}
a(t) \propto t^{\Big(-1+\sqrt{1+\frac{8\pi G_0}{3}{{t_0}^2}\rho_{v}+\frac{2}{3}\omega} \Big)}
\end{eqnarray}

 where equations (3) and (4) give the expressions for $G$ and scale factor $a(t)$.
\section{Accretion of Vacuum energy by PBH}

PBH mass can be changed by accumulating vacuum energy. The mass of PBH increases due to accretion with a rate given by

\begin{equation}
{\dot{M}}_{acc}(t)=4\pi f_{vac} {R_{BH}}^2\rho_v
\end{equation} 

where $R_{BH}=2GM$ is the black hole radius, $\rho_v={\Omega_{\Lambda}}^0\rho_{critical}$ and $f_{vac}$ denotes the vacuum energy accretion efficiency.
Now using the above expressions of $R_{BH}$ and $\rho_v$ with equation (3), we can write equation (6) as

\begin{equation}
{\dot{M}}_{acc}(t)=16\pi G_0^2\Big(\frac{t_0}{t}\Big)^4 f_{vac}M_{acc}^2\Omega_{\Lambda}^0\rho_{critical}
\end{equation}

On integration the above equation gives

\begin{eqnarray}
{\dot{M}}_{acc}(t)=M_i{\Big[1+\frac{16}{3}\pi G_0^2 t_0^4 f_{vac}\Omega_{\Lambda}^0 \rho_{critical} {M_i} (t^{-3}-t_i^{-3}) \Big]}^{-1}
\end{eqnarray}
where $M_i$ is an initial mass of PBH formed at time $t_i$ i.e $M_i=G^{-1}t_i=\Big[G_0\Big(\frac{t_0}{t_i}\Big)^2 \Big]^{-1}t_i$. 

Now the equation (8) becomes
\begin{eqnarray}
{\dot{M}}_{acc}(t)=M_i{\Big[1+\frac{16}{3}\pi G_0 t_0^2 \Omega_{\Lambda}^0 \rho_{critical}f_{vac} \Big\{\Big(\frac{t_i}{t}\Big)^3 -1\Big\} \Big]}^{-1}
\end{eqnarray}

Now using the numerical values of different quantities like $G_0=6.67 \times 10^{-8}dyne-cm^2/g^2$, $\rho_{critical}=1.1 \times 10^{-11}g/cc$, $t_0=4.42 \times 10^{17} s$ and $\Omega_{\Lambda}^0 = 0.683$, we obtain

\begin{equation}
M_{acc}(t) = M_i\Big[1 + 1.639 f_{vac} \Big\{\Big(\frac{t_i}{t}\Big)^3-1\Big\}\Big]^{-1}
\end{equation}

This equation gives the mass of PBH due to accretion only.

We can also write equation (10) as
\begin{equation}
M(t) = M_i{\Big[1-1.639 f_{vac}\Big\{1-\Big(\frac{t_i}{t}\Big)^3\Big\} \Big]}^{-1}
\end{equation}

For validity of above equation
\begin{equation}
f_{vac} < \frac{1}{1.639 \Big\{1-(\frac{t_i}{t})^3\Big\}}
\end{equation}

For large time t, $t_i/t  \to 0$ and we get
\begin{equation}
f_{vac} < \frac{1}{1.639}\simeq 0.61  
\end{equation}
The variation of PBH mass with time due to accretion only is shown in figure-1.
\begin{figure}[h]
\centering
\includegraphics[scale=0.4]{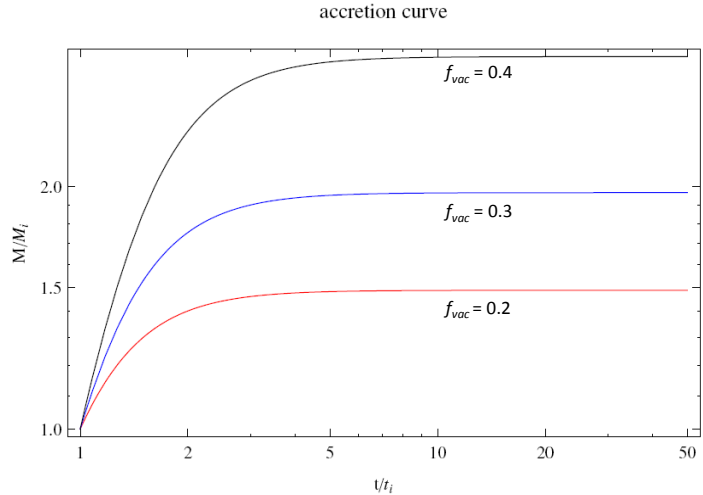}
\caption{Variation of PBH mass with time for different accretion efficiency.}
\end{figure}

The figure-1 shows that at a particular time the mass of PBH increases with increase in accretion efficiency and for a particular accretion efficiency mass saturates after a particular time.  

\section{PBH Evaporation in Vacuum Dominated Era}

Due to Hawking radiation \cite{hawk}, the rate of decrease of PBH mass is given by

\begin{equation}
\dot{M}_{evp} = -4\pi R_{BH}^2 a_H T_{BH}^4
\end{equation}
 
where $a_{BH}$ denotes the black body constant, $T_{BH}=\frac{1}{8{\pi}GM}$ is the Hawking temperature.

Using the solution of $G(t)$ with the above expressions of  $R_{BH}$ and $T_{BH}$ in vacuum dominated era we can write the rate of decrease of PBH mass as 

\begin{equation}
\dot{M}_{evap}=-\frac{a_H}{256\pi ^3} \frac{1}{G_0^2(\frac{t_0}{t})^4 M^2}
\end{equation}

We, now, study the evolution of PBH by considering both accretion and evaporation in vacuum dominated era.

In this case, the rate of change of PBH mass is given by

\begin{equation}
\dot{M}_{PBH}(t)=16{\pi}G_0^2\Big(\frac{t_0}{t}\Big)^4f_{vac}\Omega_{\Lambda}^0\rho_{critical}M_{PBH}^2-\frac{a_H}{256\pi^3}\frac{1}{G_0^2(\frac{t_0}{t})^4 M_{PBH}^2}
\end{equation}
This equation can not be solved analytically. So we use numerical method to solve it.

The evaporation of PBH for different accretion efficiencies is shown in figure-2.
\begin{figure}[h]
\centering
\includegraphics[scale=0.4]{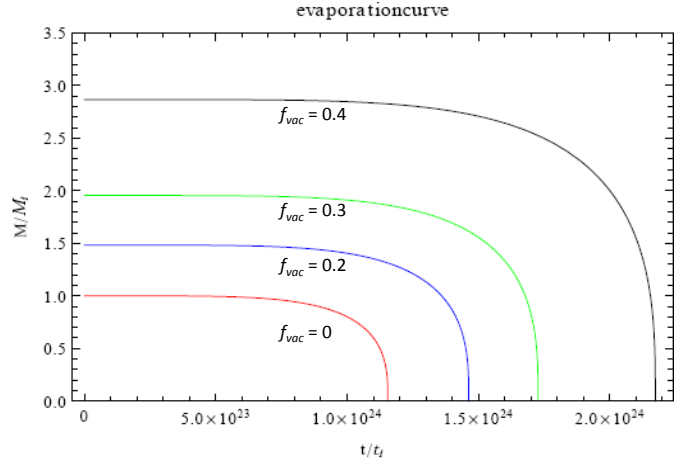}
\caption{Evaporation of a PBH having formation time $t_i=2.21\times 10^{17}s$ for different accretion efficiencies }
\label{fig2} 
\end{figure}

The figure - 2 shows that with increase in accretion efficiency life-time of PBH increases.

\section{Comparision with Standard Cosmology}

Comparision of the above results with those of Standard Cosmology [1] shows many interesting differences.
In Standard Cosmology vacuum energy accretion is only possible upto a critical time $t_c$ which depends on accretion efficiency and formation time and beyond which accretion stops while in Brans-Dicke theory vacuum energy accretion is possible for all values of time (t) eventhough the accreting mass saturates beyond certain time depending on accretion efficiency. We also observe that in BD theory the accretion of vacuum energy by PBH occurs at a smaller rate leading to quicker evaporation compared with Standard Cosmology \cite{nj}. Further, the evaporation time in BD theory is found to depend upon vacuum energy accretion efficiency whereas in Standard Cosmology it is independent of the accretion efficiency.

The evaporation time of PBH formed at $t_i=2.21 \times 10^{17}s$ for  both Brans-Dicke (BD) theory and Standard Cosmology (SC) are shown in the table-1.
\begin{table}[h]
\centering
\begin{tabular}[c]{|c|c|c|}
\hline
\multicolumn{3}{|c|}{${t_i =2.21\times10^{17}s}$}\\
\hline
$f$ & $(t_{evap})_{BD}$ & $(t_{evap})_{SC}$ \\
\hline
0 & $2.547\times10^{41}s$ & $3.695\times10^{59}s$\\
\hline
0.2 & $3.233\times10^{41}s$ & $1.867\times10^{71}s$\\
\hline
0.4 & $4.831\times10^{41}s$ & $1.867\times10^{71}s$\\
\hline
0.6 & $3.012\times10^{42}s$ & $1.867\times10^{71}s$\\
\hline
\end{tabular}
\caption{ Evaporation times of PBHs which are formed at starting of vacuum-dominated era  for different accretion efficiencies.}
\end{table}

\section{Conclusion}
Here we study the evolution of primordial black hole (PBH) in vacuum-dominated era within Brans-Dicke cosmology. First we obtain the expression for scale factor $a(t)$ and the gravitational constant $G(t)$ in vacuum-dominated era. Then we study the accretion of vacuum energy by PBHs. First we notice that the accretion efficiency has an upperbound of $0.61$. We, then, find that PBH mass increases with accretion efficiency leading to longer life as the accretion efficiency takes higher values. 

A comparision with results of Standard Cosmology \cite{nj} reveals that life-time of PBHs are many orders of magnitude smaller as a result of slow accretion rate.

\end{document}